\documentclass[12pt,a4paper]{article}
\usepackage{amssymb}
\usepackage[T2A]{fontenc}
\usepackage[cp866]{inputenc}
\usepackage[russian,english]{babel}
\usepackage{graphicx}
\usepackage{psfrag}
\usepackage{hhline}

\usepackage[unicode]{hyperref}

\begin{document}
\def\risheight{10cm}
\def\riswidth{6cm}
\begin{center}
{\Large Could the static properties of nuclei be deduced \\[2mm]
from the dynamics of a single quark?}\\[3mm]
{B.~P.~Kosyakov${}^{1,2\ast}$, E.~Yu.~Popov${}^1$, and  M. A. Vronski\u{\i}${}^{1,3}$}\\[3mm]
{{\small ${}^1$Russian Federal Nuclear Center--VNIIEF, 
Sarov, 607190 Nizhni\u{\i} Novgorod Region, Russia;\\
${}^2$Moscow Institute of Physics {\&} Technology, Dolgoprudny\u{\i}, 141700 Moscow Region; 
Russia; 
${}^3$Sarov Institute of Physics {\&} Technology, Sarov, 607190 Nizhni\u{\i} Novgorod Region, Russia.}\\
{\tt E-mail address:} 
${\rm  kosyakov.boris@gmail.com}$} 
\end{center}
\begin{abstract}
\noindent
{
We show that the static properties of a nucleus could arise from a single quark 
moving in a mean field generated by all other constituents of this nucleus.
The resulting model provides a way for determining the nuclear sizes characteristic 
of the liquid drop model, and reasonably accurate values of magnetic moments of 
different nuclei with the aid of two parameters $\alpha_s$ and $\sigma$ 
appearing in the Cornell potential intended for use in quarkonium physics.
}
\end{abstract}

\vskip3mm

\vskip15mm

\noindent
Shortened version of the title:

\vskip5mm

{\Large Static properties of nuclei from the dynamics of a single quark}

\vskip80mm
${}^{\ast}$ Corresponding author

\newpage
\section{Introduction}
\label
{Introduction}
With the advent of quantum chromodynamics (QCD) a serious effort was mounted to 
understand nuclei in terms of quarks.
Early in the development of this line of inquiry, a nucleus with mass number 
${\cal A}$ was conceived as a system of $N=3{\cal A}$ quarks moving in a large bag 
\cite{Bleuler}, \cite{Petry}.
However, the relationship between the number of quarks that are contained 
in a stable bag and its size $R$ is given\footnote{See, e.~g., Ref.~\cite{Close}, 
Eq.~(18.20).} by $R\sim N^{1/4}$ which is contrary to the experimental nuclear 
data.
Furthermore, the predicted nuclear magnetic moments significantly differ from 
their established values \cite{Arima}, \cite{Talmi}.
We thus should proceed from another paradigm.  

A direct way for clarifying the properties of a many-quark system is to use the 
Feynman path integral machinery whereby all degrees of freedom are 
integrated out, except for those of a single quark $Q$, so that this quark is 
affected by a mean field generated by all other constituents of the system.
A systematic implementation of this calculation program is still a good distance 
in the future\footnote{An effort to integrate out only gluon degrees of freedom 
was reasonably successful \cite{Maltman}, 
but never progressed beyond small nuclei.}.
And yet the result of integrating out the remaining degrees of freedom can be
neatly approximated if we make plausible assumptions about the mean field, and 
invoke the semiclassical approach for exploring the behavior of the quark $Q$.
Is the semiclassical approach suitable to nuclear physics?
It is pertinent to note that some properties of nuclei are typical of classical 
entities.
For example, the observed size of a nucleus $R$ depends on mass 
number ${\cal A}$ as 
\begin{equation}
{R}={R_0}\, {\cal A}^{1/3}\,,
\label
{R-nucleus}
\end{equation}                                           
which is peculiar to a classical liquid drop rather than a quantum object whose 
extension, characterized by its Compton wavelength, is inversely proportional
to  ${\cal A}$.

An attempt to develop these ideas showed considerable promise \cite{KPV}. 
Some properties of nuclei are indeed amenable to this treatment.
The nuclear force saturation (the property which became pressing in  
QCD-inspired approaches), and the nuclear scale characteristic of the liquid drop 
model, Eq.~(\ref{R-nucleus}), are notable examples.
We will show in the present paper that reasonably accurate values of 
magnetic moments for a rich variety of nuclei are also attainable in this 
analysis.

The dynamics of the quark $Q$, specified by the Dirac field $\psi$, is assumed 
to be encoded by the Lagrangian 
\begin{equation}
{\cal L}
={\bar\psi}\left[i\gamma^\mu\left(\partial_\mu+ig_VA_\mu\right)-m_Q\right]\psi 
+ g_S{\bar\psi}\psi\,\Phi\,,
\label
{QCD-Lagrangian}
\end{equation}                                           
where $A_\mu=(A_0,{\bf A})$ and $\Phi$ are respectively the Lorentz vector
potential and Lorentz scalar potential of the mean field, $g_V$ and $g_S$ 
their associated couplings, and $m_Q$ the current-quark mass of the quark $Q$. 
The second interaction term in (\ref{QCD-Lagrangian}) is absent from the 
QCD Lagrangian because the scalar Yukawa coupling is contrary to 
asymptotic freedom.
However, our concern is with the effective theory in the infrared region 
where the dynamics is anticipated to arrange itself into the form shown in 
Eq.~(\ref{QCD-Lagrangian}).

The semiclassical treatment implies that the extremal path contribution 
dominates the Feynman path integral.
This is the same as saying the wave function $\psi(x)$ of the quark $Q$ is a 
solution to the Dirac equation in the background $A_\mu(x)$ and $\Phi(x)$ because this equation 
results from the requirement that the action with the Lagrangian (\ref{QCD-Lagrangian}) 
be extremal.
In fact, we restrict our attention to spherically symmetric static interactions, 
and assume that the contribution of the Lorentz vector potential to the mean
field is given by $A_0$, that is, ${\bf A}=0$. 
What this means is a particle with reduced mass $m$ orbits the center of mass 
being driven by central potentials  $A_0(r)$ and $\Phi(r)$.
The assumption of the spherically symmetric interaction as applied to the case
that the quark $Q$ moves in the mean field of  an intricate form may seem 
awkward.
However, the arguments in support of this assumption closely resemble those 
taken in the single-particle shell model of atomic nuclei in which the mean 
field exerting on 
every nucleon is given by a central potential because the nucleus in its ground 
state is approximately spherically symmetric, and the Pauli principle, 
acting through the already occupied orbitals, suppresses the role of 
long-range correlations \cite{Walecka}.
We thus proceed from the Dirac Hamiltonian 
\begin{equation}
H=-i{\bf\alpha}\cdot{\nabla}+U_V({r})+\beta [m+U_S({r})]\,,
\label
{Dirac-Hamiltonian}
\end{equation}
where $U_V=g_VA_0$, and $U_S=g_S\Phi$. 

It will be recalled that the Dirac equation can be safely regarded as a 
one-particle wave equation for interactions of a special kind.
Following the conventional interpretation, positive energy states 
are attributed to a Dirac particle, while states of negative energy are 
attributed to its antiparticle.
If there is a unitary transformation which diagonalizes the Dirac Hamiltonian
with respect to positive and negative energies, then the wave functions of a 
Dirac particle of definite momentum have just two components, as it must for 
the usual quantum-mechanical interpretation of these wave functions to be 
adequate.
Foldy and Wouthuysen \cite{Foldy--Wouthuysen} found a unitary transformation 
which diagonalizes the free Dirac Hamiltonian with respect to positive and 
negative energies.
Case \cite{Case} obtained a closed form for this transformation in the presence 
of a time-independent magnetic field of arbitrary strength.
The energy gap of the Dirac sea is not penetrated in the constant magnetic field because this 
field leaves the energy of the Dirac particle unchanged.  
Note also that the block-diagonalization is another way of stating that the system enjoys the 
property of supersymmetry \cite{Zentella}. 
The pseudoscalar coupling is a further fascinating candidate for the 
block-diagonalization of the Dirac Hamiltonian.
Indeed, adding the term $g_P{\bar\psi}\gamma_5\psi\,\chi$ to (\ref{QCD-Lagrangian}),
and setting $g_V=g_S=0$, we come to the Foldy--Wouthuysen picture.
The systems with pseudoscalar couplings will be analyzed elsewhere.

Our prime interest in the present paper is with the systems which are governed 
by the Hamiltonian defined in (\ref{Dirac-Hamiltonian}) and subject to the 
spin- and pseudospin symmetry conditions (for an 
extended discussion of these  phenomenological symmetries see Refs.~\cite{Ginocchio} and 
\cite{Liang}).
The Foldy--Wouthuysen separation of positive- and negative-energy states is not
attained in such  systems.
We will ascribe the spin symmetry condition to free hadrons, and the pseudospin 
symmetry condition to nuclei.

We take the Cornell potential \cite{Cornell75} 
\begin{equation}
{V_{\rm C}(r)}=
-\frac{\alpha_s}{r}+\sigma r \,
\label
{Cornell}
\end{equation}
as a phenomenological realization of both $U_V$ and $U_S$, and put 
$U_V=\frac12\,V_{\rm C}$.
Could the potential (\ref{Cornell}) be as much useful in nuclear physics as in 
the research of quarkonia?
It will transpire in Sect.~3 that the pseudospin symmetry condition gives 
rise to a nucleus-sized cavity with singular boundary capable of keeping the
quark $Q$ in this cavity from escaping whenever $U_V(r)$ grows indefinitely with $r$. 
We are thus free to vary rising potentials $U_V(r)$ and $U_S(r)$ in a wide 
range of their forms to attain the best fit to experiment.
In the present paper, the form of $|U_V|$ and $|U_S|$ is fixed to be identical to 
that of $\frac12\,|V_{\rm C}|$, and the values of $\alpha_s$ and $\sigma$ are 
borrowed from the description of quarkonia, -- to examine what results
in nuclear physics.

We are now in position to formulate our strategy.
We assume that an effective theory to low energy QCD, which covers both free 
hadrons and nuclei, may arise when all degrees of freedom are integrated 
out, except for those of a single quark $Q$.
To take the last step, that is, find the Feynman path integral over  
degrees of freedom of the quark $Q$, we would have to substitute this procedure 
with its semiclassical approximation.
This leads us to the eigenvalue problem  
\begin{equation}
\left\{-i{\bf\alpha}\cdot{\nabla}+
U_V({r})+\beta [m+U_S({r})]
\right\}\psi({\bf r})=\varepsilon\psi({\bf r})\,.
\label
{Dirac-genera}
\end{equation}
We look for solutions of this problem combined with the spin- and pseudospin 
symmetry conditions.
We separate variables in the two-row Dirac eigenstates, 
\begin{equation}
\psi_{n_r,\ell,j,M}({\bf r})
=
\pmatrix{f_{n_r,\ell}(r)\left[Y^{(\ell)}(\theta,\phi)\chi\right]^{(j)}_M\cr
         ig_{n_r,\ell,j}(r)\left[Y^{(\ell_j)}(\theta,\phi)\chi\right]^{(j)}_M\cr}.
\label
{Dirac-eigenstate}
\end{equation}
Here, $f_{n_r,\ell}$ and $g_{n_r,\ell,j}$ are the radial amplitudes, $n_r$, 
$\ell$, $j$ are the radial, orbital, and total angular momentum quantum 
numbers, respectively, $\left[Y^{(\ell)}(\theta,\phi)\chi\right]^{(j)}_M$ stands 
for the coupled amplitude 
$\sum_{m,\mu}C^{\ell(1/2)j}_{m \mu M}Y^{(\ell)}(\theta,\phi)\chi_\mu$,
$Y^{(\ell)}(\theta,\phi)$ is the spherical harmonic of order $\ell$, 
$\ell_j$ is given by $\ell_{\ell-1/2}=\ell-1$ and $\ell_{\ell+1/2}=\ell+1$, 
$\chi_\mu$ is the spin function.
The operator ${K}=-\beta\,({\bf S}\cdot{\bf L}+1)$
commutes with the spherically symmetric Dirac Hamiltonian.
Thus, the Dirac eigenstates are the eigenstates of this operator with 
eigenvalues $\kappa=\pm(j+\frac12)$, with `$-$' for aligned spin 
$(s_{1/2},p_{3/2},{\rm etc}.)$, and `$+$' for unaligned spin 
$(p_{1/2},d_{3/2},{\rm etc}.)$, so that the quantum number
$\kappa$ is sufficient to label the orbitals \cite{Ginocchio-}.
The radial part of Eq.~(\ref{Dirac-genera}) is
\begin{equation}
{f'}+\frac{1+\kappa}{r}\,f-ag=0\,,
\label
{Dirac_radia_f}
\end{equation}
\begin{equation}
{g'}+\frac{1-\kappa}{r}\,g+bf=0\,,
\label
{Dirac_radia}
\end{equation}
\begin{equation}
a(r)=\varepsilon+{m}+U_S(r)-U_V(r)\,,
\label
{A-df}
\end{equation}
\begin{equation}
b(r)=\varepsilon-{m}-U_S(r)-U_V(r)\,,
\label
{B-df}
\end{equation}
where the prime stands for differentiation with respect to $r$.

We use (\ref{Dirac_radia_f}) for expressing $g$ in terms of $f$ and 
substitute the result in (\ref{Dirac_radia})\footnote{Our concern is with $f$ 
because it is $f$ that survives in the nonrelativistic free-particle limit.}.
We eliminate the first derivative of $f$ from the resulting second-order 
differential equation 
\begin{equation}
f''+Af'+Bf=0\,,
\label
{second-order-eq}
\end{equation}                                          
in which
\begin{equation}
A=-\frac{a'}{a}+\frac{2}{r}\,,
\quad 
B=a\,({1+\kappa})\left(\frac{1}{{r}a}\right)'+ab+\frac{1-\kappa^2}{r^2}\,,
\label
{B_1-df}
\end{equation}
using the ansatz 
\begin{equation}
f=F\,\frac{\sqrt{a}}{r}\,,
\label
{ansatz}
\end{equation}                                          
to obtain the Schr\"odinger-like equation
\begin{equation}
F''+k^2F=0\,,
\label
{1D_Schroedinger}
\end{equation}                                          
where
\begin{equation}
k^2={\varepsilon^2-m^2}-2U(r;\varepsilon)=-\frac12\,A'(r)-\frac14\,A^2(r)
+B(r)\,\,.
\label
{k-df}
\end{equation}

Once all angular (orbital and spin) variables are thus eliminated, and the
pertinent phenomenological condition on $U_S$ and $U_V$ is imposed,  
the function $U(r;\varepsilon)$ defined in (\ref{k-df}) 
acts as the effective potential.

The Schr\"odinger-like equation (\ref{1D_Schroedinger}) is formally equivalent 
to the set of equations (\ref{Dirac_radia_f})--(\ref{Dirac_radia})  \cite{Liang--}.
However, the set of equations (\ref{Dirac_radia_f})--(\ref{Dirac_radia}) is not 
diagonalizable, which seems invalidate the probabilistic interpretation of the two-row wave function 
of the Dirac particle $Q$.
Meanwhile the particle $Q$ is not an ordinary quantum-mechanical particle 
because no quark can be isolated.
From the conceptual point of view, the quark $Q$ defies its probing everywhere
outside the region to which this quark is confined.
We therefore will assume that $F$, the solution to the one-dimensional 
Schr\"odinger-like equation (\ref{1D_Schroedinger}), is just the 
probability amplitude of the quark $Q$.

\section{Quarkonia}
Spin symmetry is inherent in free hadron states \cite{Goldman}, \cite{Ginocchio--}.
This symmetry occurs when $U_S=U_V$.
With $V_{\rm C}=2U_V$, Eqs.~(\ref{k-df}) and (\ref{B_1-df}) can be solved to 
give the effective potential
\begin{equation}
U(r;\varepsilon)=\frac{1}{2}\left[\frac{\kappa(\kappa+1)}{r^2}
+\left(\varepsilon +{m}\right)\left(-\frac{\alpha_s}{r}
+\sigma r\right)\right].
\label
{U_eff}
\end{equation}

In the nonrelativistic limit $\varepsilon\to m$,  $U(r;\varepsilon)$
becomes the sum of the centrifugal term and the Cornell potential, and hence, 
the results for the spectrum of quarkonia which were obtained through the use 
of the Schr\"odinger equation \cite{Cornell80}, \cite{Barnes} are reproduced in 
this procedure.
It is then clear that spin-dependent effects appear as small corrections.

The procedure of looking for numerical solutions to Eqs.~(\ref{Dirac_radia_f})--(\ref{B-df})
with imposing the conditions $U_S = U_V = \frac12\,V_C$ is outlined in 
Appendix.
The agreement between the energy levels $\varepsilon_{n_r}$ obtained in this 
procedure and the masses of well established $c{\bar c}$ and $b{\bar b}$ states is 
within $\sim 1\%$ for $\alpha_s=0.7$, $\sigma=0.14\,{\rm GeV}^2$, $m_c=1.45$ GeV, 
$m_b=4.92$ GeV, see Table~\ref{tabone} which also lists the experimental data from 
\cite{Olive}. 

A precise calculation of binary system energy levels should 
proceed from the use of the Bethe--Salpeter equation. 
However, both  $c{\bar c}$ and $b{\bar b}$ are nonrelativistic systems, so that 
reasonably accurate results for the spectrum (which is experimentally measured 
to within $\sim 1\%$) can be found by a nonrelativistic 
machinery, the Schr\"odinger equation governing the behavior of a particle with 
reduced mass $m=\frac12\,m_Q$ in static confinement potentials, such as the 
Cornell potential \cite{Cornell80-, Barnes-}.
The use of the Dirac equation for this purpose is as good as that of
the Schr\"odinger equation.
\begin{table}
\caption{Quarkonium masses}\label{tabone}
\begin{center}
\begin{tabular}{|c|c|c|c|c|c|}\hline
$M_{c\overline c}$&Calculation      &   Experiment    &$M_{b\overline b}$ &Calculation    & Experiment\\\hline
$J/\psi(1S)$      &     3.084       &   3.097    &    $\gamma(1S)$        &   9.415       &   9.460   \\\hline
$\eta_c(1S)$      &     3.084       &   2.980    &    $\eta_b(1S)$        &   9.415       &   9.398   \\\hline
$\psi(2S)$        &     3.635       &   3.686    &    $\gamma(2S)$        &   10.04       &   10.023  \\\hline
$\eta_c(2S)$      &     3.635       &   3.638    &    $\eta_b(2S)$        &   10.04       &   9.999   \\\hline
$\psi(3S)$        &     4           &   4.030    &    $\gamma(3S)$        &   10.35       &   10.355  \\\hline
$\psi(4S)$        &     4.3         &   4.421    &    $\gamma(4S)$        &   10.582      &   10.579  \\\hline
$\chi_{c0}(1P)$   &     3.511       &   3.414    &    $\gamma(5S)$        &   10.777      &   10.865  \\\hline
$\chi_{c1}(1P)$   &     3.511       &   3.510    &    $\gamma(6S)$        &   10.948      &   11.019  \\\hline
$\chi_{c2}(1P)$   &     3.511       &   3.556    &    $\chi_{b0}(1P)$     &   9.98        &   9.859   \\\hline
$h_c(1P)$         &     3.511       &   3.525    &    $\chi_{b1}(1P)$     &   9.98        &   9.892   \\\hline
$\chi_{c2}(2P)$   &     3.897       &   3.929    &    $\chi_{b2}(1P)$     &   9.98        &   9.912   \\\hline
$\psi(1D)$        &     3.773       &   3.771    &    $\chi_{b0}(2P)$     &   10.297      &   10.232  \\\hline
$\psi(2D)$        &     4.095       &   4.153    &    $\chi_{b1}(2P)$     &   10.297      &   10.255  \\\hline
                  &                 &            &    $\chi_{b2}(2P)$     &   10.297      &   10.268  \\\hline
                  &                 &            &    $1^3D_2$            &   10.222      &   10.161  \\\hline
\end{tabular}
\end{center}
\end{table} 

\section{Nuclei}
The wavefunction classification stemming from pseudospin symmetry is important 
for both light and very heavy nuclei whose 
superdeformation appears already at low spin \cite{Ginocchio---}, 
\cite{Liang-}.
The original observations that lead to the concept of ``pseudospin symmetry'' 
were quasi-degeneracies in spherical shell model orbitals with 
$(n_r, \ell, j=\ell+1/2)$ and $(n_r-1, \ell+2, j=\ell+3/2)$ where $n_r$, $\ell$, 
and $j$ are single-nucleon radial, orbital, and total orbital angular momentum
quantum numbers, respectively. 
Pseudospin degeneracy in heavy nuclei derives from the fact that nucleons
in a nucleus move in an attractive scalar,  $-U_S$, and
repulsive vector,  $U_V$, mean fields, which are
nearly equal in magnitude,  $|U_S|\approx |U_V|$.
This near equality of mean fields is likely a general feature of
any relativistic model which fits nuclear binding energies \cite{GinocchioPRL}.

One can go a step further and consider this condition as that arising from a 
quark $Q$ moving in a mean field generated by other quarks in the nucleus 
\cite{KPV-}.
We adopt the condition $U_S=-U_V+C_s$ where $C_s$ is a constant.
The Hamiltonian governing the quark $Q$ becomes 
\begin{equation}
H_s={\bf\alpha}\cdot{\bf p}+U_V({r})(1-\beta)+ \beta(m+C_s)\,.
\label
{Dirac-Hamiltonian-spin}
\end{equation}
We thus see that $m$ is shifted, $m\to m_s=m+C_s$. 
This shift may signal that the current-quark masses become the 
corresponding constituent-quark masses.
In what follows $m_s$ will be regarded as the constituent-quark mass of the quark 
$Q$, and the label $s$ of $m_s$ will be omitted.

We put $V_{\rm C}=2U_V$,  and solve Eqs.~(\ref{k-df}) and (\ref{B_1-df})
to give    
\[
U(r;\varepsilon)=\frac{1}{2{r^2}}\Biggl\{{\kappa(\kappa+1)}
+\left(\varepsilon-{m}\right)\left(-\frac{\alpha_s}{r}
+\sigma r\right){r^2}
\]
\begin{equation}
+\frac{3(\alpha_s+\sigma r^2)^2}
{4\left[\sigma r^2-\left(\varepsilon +{m}\right)r-\alpha_s\right]^2}+
\frac{\alpha_s(\kappa+1)+\kappa\sigma r^2}
{\sigma r^2-\left(\varepsilon +{m}\right)r-\alpha_s}
\Biggr\}\,.
\label
{U_eff-PSEUDO}
\end{equation}

\begin{figure}[htb]
\centerline{\includegraphics[height=\risheight,angle=270]{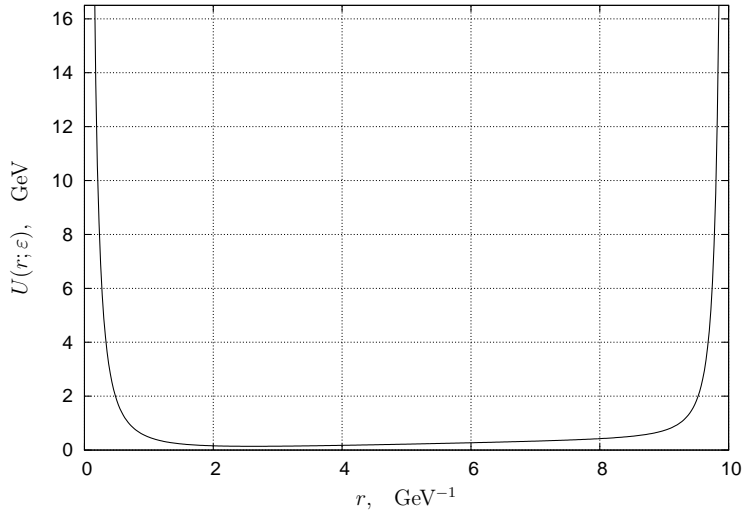}}
\caption{The effective potential (\ref{U_eff-PSEUDO}) with the parameters
$m=0.33\,{\rm GeV}$, $\varepsilon=1\,{\rm GeV}$, 
$\alpha_s=0.7$, $\sigma=0.14\,{\rm GeV}^2$, $\kappa=1$}\label{pseudo-spin-potential}
\end{figure}

The terms in the first line of (\ref{U_eff-PSEUDO}) closely resemble the 
respective terms involved in (\ref{U_eff}), except for changing the overall 
factor of the Cornell potential, but the terms of the second line dramatically 
change the situation.
They are singular at $r=r_{\rm sc}$ which is
the positive root of the equation $\sigma r^2-\left(\varepsilon +{m}\right)r-\alpha_s=0$,
\begin{equation}
r_{\rm sc}=\frac{\left(\varepsilon +{m}\right)+
\sqrt{\left(\varepsilon +{m}\right)^2+4\sigma \alpha_s}}{2\sigma}\,.
\label
{sol}
\end{equation}
To illustrate, the form of the effective potential (\ref{U_eff-PSEUDO}) with 
some specific values of $m$, $\varepsilon$, $\alpha_s$, $\sigma$, and $\kappa$ 
is depicted in Fig.~\ref{pseudo-spin-potential}.

\psfrag{x}[c][c][0.7]{$r$,\quad GeV$^{-1}$}
\psfrag{y}[c][c][0.7]{$U(r;\varepsilon)$,\quad GeV}

It is thus seen that the condition $U_V=-U_S=\frac12\,V_{\rm C}$ vastly enhances 
the interaction between the quark $Q$ and the mean field to yield a spherical 
shell of radius $r_{\rm sc}$ on which $U(r;\varepsilon)$ is infinite. 
The boundary of the spherical cavity of radius $r_{\rm sc}$ keeps the quark $Q$ 
in this cavity from escaping.
It is well known \cite{Dittrich} that the tunneling through a potential barrier 
of the form $\lambda(x-x_0)^{-2}$  with $\lambda\ge \frac34$
is forbidden in one-dimensional quantum mechanics.
This condition is fulfilled by Eq.~(\ref{U_eff-PSEUDO}).
Therefore, the boundary of the cavity sets up an impenetrable quantum-mechanical 
barrier.

A singular boundary arises whenever $U_V(r)$ grows indefinitely with $r$. 
This is because in going from Eqs.~(\ref{Dirac_radia_f}) and (\ref{Dirac_radia}) to 
Eq.~(\ref{1D_Schroedinger}), we have to apply the factor $1/a$ which is infinite 
when $a=0$.
Such is not the case when we adopt the spin symmetry condition $U_S=U_V$  
by which $a=\varepsilon+{m}$.
In contrast, the pseudospin symmetry condition $U_S=-U_V$ implies that 
$a=\varepsilon+{m}-2U_V$, and $a=0$ has a positive root provided that $U_V$ 
increases monotonically with $r$ beginning at $r=0$ where $U_V$ assumes a 
negative value.
However, no singular boundary arises when $U_V\to U_0$ as 
$r\to\infty$, where $U_0$ is a constant which is less than 
$\frac12\left(\varepsilon+{m}\right)$. 
This is the reason for the absence of confinement from systems with 
electromagnetic bindings.

It may be worth pointing out that the mathematical problem associated with 
Eq.~(\ref{1D_Schroedinger}) is not identical to the conventional eigenvalue 
problem for a one-dimensional Schr\"odinger Hamiltonian.
Indeed, the effective potential $U(r;\varepsilon)$, defined in 
(\ref{k-df}), depends nonlinearly on the parameter $\varepsilon$ 
which plays the role of an ``eigenvalue''\footnote{This raises the question of 
how to determine the radial quantum number ${n_r}$ in this context. 
Since ${n_r}$ equals the number of nodes of the radial part of the eigenfunction 
in the conventional eigenvalue problem, we take this fact as a way for 
determining ${n_r}$ by counting the nodes of $f_{n_r,\kappa}$ or  $g_{n_r,\kappa}$.}. 
In contrast to the conventional eigenvalue problem where all energy levels are 
referred to a fixed effective potential $U(r)$, for every energy level 
$\varepsilon$ appearing in the present problem there is a unique potential 
configuration $U(r;\varepsilon)$ exhibiting the spherical cavity of radius 
$r_{\rm sc}(\varepsilon)$ peculiar to just this $\varepsilon$\footnote{Note that the 
transition of the quark $Q$ from a given state specified by some ${n_r}$ to 
another state with lesser ${n_r}$ is forbidden because this transition would 
decrease the range of its localization (the impenetrable cavity would be smaller), with both 
energy levels being assumed to be definite, which is contrary to the Heisenberg's 
uncertainty principle.}. 
With this remark in mind it is little wonder that the behavior of the quark $Q$ 
in the state whose energy $\varepsilon$ is much greater than the 
constituent-quark mass $m$ may well be nonrelativistic because the singular
interaction between this quark and the mean field 
of the nucleus converts the major portion of $\varepsilon$ to the mass 
content of the nucleus, expressed by the mass number ${\cal A}$, and only a 
tiny part of $\varepsilon$ is to be assigned to kinetic energy.                

To verify that the effective potential $U(r;\varepsilon)$ defined by Eq.~(\ref{U_eff-PSEUDO}) 
is indeed attributable to the description of 
nuclei, we solve numerically Eqs.~(\ref{Dirac_radia_f}) and (\ref{Dirac_radia}) 
using the parameters $\alpha_s=0.7$ and $\sigma=0.1\,{\rm GeV}^2$ (borrowed from
the description of quarkonia), and taking $m$ to be 0.33\,{GeV}.
The procedure closely parallels that outlined in Appendix.

We find the energy levels $\varepsilon_{n_r}$ for $\kappa=-1,-2$ 
and the corresponding sizes of the cavities $r_{\rm sc}({n_r})$.
To a good approximation the energy levels $\varepsilon_{n_r}$ turn out to be 
proportional to $\sqrt{n_r}$\footnote
{To offer a qualitative explanation of this relationship 
for large ${n_r}$, we recast 
Eqs.~(\ref{Dirac_radia_f}) and (\ref{Dirac_radia}) as $G''+\left[(\kappa-\kappa^2)/{r^2}+
\varepsilon\left(\varepsilon+m+{\alpha_c}/{r}-\sigma r\right)\right]G=0$,
where $G=rg$.
Since our concern is with large $r_{\rm sc}$, we may safely omit the terms 
proportional to $r^{-1}$ and $r^{-2}$.
This gives the Airy equation.
Regular solutions to this equation behave asymptotically as $\sin\left\{
{\frac23(\varepsilon\sigma)^{1/2}}\left[({\varepsilon+m})/{\sigma}- r\right]^{3/2}
+\phi_0\right\}$, which shows that the number of radial nodes  ${n_r}$ is 
estimated at 
$\left[{\frac23(\varepsilon\sigma)^{1/2}}/{\pi}\right]\left[({\varepsilon+m})/{\sigma}\right]^{3/2}$.
Therefore, $\varepsilon_{n_r}\sim\sqrt{n_r}$. }.
By assuming that  ${n_r}$ is proportional to ${\cal A}^{2/3}$, where 
${\cal A}$ is the nucleon mass number, we come to
the relationship $r_{\rm sc}=R_0{\cal A}^{1/3}$ with $R_0\approx 1$ fm for the 
chosen values of the $\alpha_s$ and $\sigma$, which is consistent with 
Eq.~(\ref{R-nucleus}), for more details see Table~2.

\begin{table}[!h]
\caption{Calculated nuclei sizes}\label{tabtwo}
\begin{center}
\begin{tabular}{|c|c|c|c|c|p{4mm}|c|c|c|c|c|}\hhline{-----~-----}
Element&${\cal A}$&$n_r$&$r_{\rm sc}$ (fm)&$R_0$ (fm)              && Element&${\cal A}$&$n_r$&$r_{\rm sc}$ (fm)&$R_0$ (fm)             \\\hhline{-----~-----}
\multicolumn{5}{|l|}{Neutron odd nuclei ($j=1/2$, $\varkappa=-1$)} && \multicolumn{5}{|l|}{Neutron odd nuclei ($j=3/2$, $\varkappa=-2$)}\\\hhline{-----~-----}
$_6^{13}C$&13&6&2.84&1.21                                          && $_{6}^{11}C$&11&5&2.76&1.24                                       \\\hhline{-----~-----}
$_8^{15}O$&15&6&2.84&1.15                                          && $_{16}^{33}S$&33&10&3.69&1.15                                     \\\hhline{-----~-----}
$_{16}^{31}S$&31&10&3.6&1.14                                       && $_{16}^{35}S$&35&11&3.87&1.16                                     \\\hhline{-----~-----}
$_{32}^{71}Ge$&71&17&4.65&1.13                                     && $_{18}^{35}Ar$&35&11&3.87&1.16                                    \\\hhline{-----~-----}
$_{32}^{75}Ge$&75&18&4.8&1.14                                      && $_{18}^{37}Ar$&37&11&3.87&1.14                                    \\\hhline{-----~-----}
$_{34}^{77}Se$&77&18&4.8&1.13                                      && $_{20}^{39}Ca$&39&12&4.02&1.18                                    \\\hhline{-----~-----}
$_{50}^{113}Sn$&113&23&5.42&1.12                                   && $_{28}^{57}Ni$&57&15&4.46&1.16                                    \\\hhline{-----~-----}
$_{52}^{119}Te$&119&24&5.53&1.13                                   && $_{50}^{121}Sn$&121&24&5.6&1.13                                   \\\hhline{-----~-----}
$_{68}^{169}Er$&169&31&6.29&1.14                                   && $_{52}^{129}Te$&129&26&5.82&1.15                                  \\\hhline{-----~-----}
$_{70}^{171}Yb$&171&31&6.29&1.13                                   && $_{54}^{131}Xe$&131&26&5.82&1.15                                  \\\hhline{-----~-----}
$_{80}^{181}Hg$&181&32&6.4&1.13                                    && $_{56}^{135}Ba$&135&26&5.82&1.14                                  \\\hhline{-----~-----}
$_{80}^{183}Hg$&183&32&6.4&1.12                                    && $_{56}^{137}Ba$&137&27&5.93&1.15                                  \\\hhline{-----~-----}
$_{80}^{185}Hg$&185&32&6.4&1.12                                    && $_{58}^{137}Ce$&137&27&5.93&1.15                                  \\\hhline{-----~-----}
$_{78}^{195}Pt$&195&34&6.59&1.13                                   && $_{58}^{139}Ce$&139&27&5.93&1.15                                  \\\hhline{-----~-----}
$_{78}^{197}Pt$&197&34&6.59&1.13                                   && $_{58}^{143}Ce$&143&27&5.93&1.13                                  \\\hhline{-----~-----}
$_{82}^{207}Pb$&207&35&6.69&1.13                                   && $_{74}^{187}W$&187&33&6.54&1.14                                   \\\hhline{-----~-----}
$_{84}^{209}Po$&209&35&6.69&1.13                                   && $_{76}^{189}Os$&189&33&6.54&1.14                                  \\\hhline{-----~-----}
\multicolumn{5}{|l|}{Proton odd nuclei ($j=1/2$, $\varkappa=-1$)}  &\multicolumn{6}{l}{}\\\hhline{-----~~~~~~}
$_9^{19}F$&19&7&3.03&1.14                                          &\multicolumn{6}{l}{}\\\hhline{-----~~~~~~}
$_{15}^{29}P$&29&9&3.43&1.12                                       &\multicolumn{6}{l}{}\\\hhline{-----~~~~~~}
$_{15}^{31}P$&31&10&3.6&1.14                                       &\multicolumn{6}{l}{}\\\hhline{-----~~~~~~}
$_{55}^{125}Cs$&125&25&5.65&1.13                                   &\multicolumn{6}{l}{}\\\hhline{-----~~~~~~}
$_{55}^{127}Cs$&127&25&5.65&1.12                                   &\multicolumn{6}{l}{}\\\hhline{-----~~~~~~}
$_{55}^{129}Cs$&129&26&5.77&1.12                                   &\multicolumn{6}{l}{}\\\hhline{-----~~~~~~}
$_{81}^{195}Tl$&195&34&6.58&1.13                                   &\multicolumn{6}{l}{}\\\hhline{-----~~~~~~}
$_{81}^{197}Tl$&197&34&6.58&1.13                                   &\multicolumn{6}{l}{}\\\hhline{-----~~~~~~}
\end{tabular}
\end{center}
\end{table}

The conjecture that ${n_r}$ equals the integral part of ${\cal A}^{2/3}$ 
has far-reaching implications, sending us in search of evidence for or against 
this conjecture. 
With this in mind, we compare the magnetic dipole of the quark $Q$ and that of the 
nucleus in which this quark is incorporated.
Note that the magnetic moment of a nucleon in
the states with isospin $I$ and the total angular momentum 
$j={\ell}+\frac12$ and $j={\ell}+\frac32$
is given  \cite{GinocchioPRC}, respectively, by 
\begin{equation}
\mu_{j,I}=-\frac{e\gamma_I\left(j+\frac12\right)}{2\left(j+1\right)}\int_0^\infty
f_{{n}_r-1,{\ell}+1,j,I}g_{{n}_r-1,{\ell}+1,I}r^3dr
+\mu_{A,I}\left(1-\frac{2j+1}{j+1}\int_0^\infty
g^2_{{n}_r-1,{\ell}+1,I}r^2dr\right),
\label
{mu-j-1_2}
\end{equation}
and
\begin{equation}
\mu_{j,I}=\frac{e\gamma_I\left(j+\frac12\right)}{2\left(j+1\right)}\int_0^\infty
f_{{n}_r-1,{\ell}+1,j,I}g_{{n}_r-1,{\ell}+1,I}r^3dr
-\frac{\mu_{A,I}}{j+1}\left[j-(2j+1)\int_0^\infty
g^2_{{n}_r-1,{\ell}+1,I}r^2dr\right].
\label
{mu-j+1_2}
\end{equation}
Here, $\gamma_I$ is the orbital gyromagnetic ratio, $\mu_{A,I}$ is the anomalous 
magnetic moment,  $\mu_{A,-\frac12}=-1.913\,\mu_{\rm B}$, $\mu_{A,\frac12}=1.793\,\mu_{\rm B}$, where 
$\mu_{\rm B}=e/2M$ is the nucleon magneton.
We reiterate mutatis mutandis the arguments of Ref.~\cite{GinocchioPRC-} to 
conclude that (\ref{mu-j-1_2}) and (\ref{mu-j+1_2}), in which the infinite
limits of integration are replaced by $r_{\rm sc}$, are well suited for
the magnetic moment of neutron-odd nuclei with 
$j= \frac12$ and $j=\frac32$ to be represented by the magnetic moment of a 
constituent $u$ 
quark contained in these nuclei, with the understanding that ${n_r}$ 
is associated with $\left[{{\cal A}}^{2/3}\right]$, where the square brackets
denote the integral part of the quantity enclosed in them.
We determine the quark wave functions $f_{{n}_r,j}$ and $g_{{n}_r,j}$ for 
$\alpha_s=0.7$, $\sigma=0.14\,{\rm GeV}^2$, $m=0.33\,{\rm GeV}$.
As to the anomalous magnetic moments of quarks, the present notion of their 
values is far from complete.
Among suggested values of $\mu_{A}$ for $u$ quarks \cite{Bicudo}, \cite{Mekhfi}, 
we adopt to test $\mu_{A}=0.15\,{\mu}_{\rm B}$, and $ 0.2\,{\mu}_{\rm B}$.

Fig.~2(a,b) shows that the magnetic moments of 
neutron-odd nuclei with $j= \frac12$ and $j=\frac32$ \cite{Fuller} agree with 
the calculated magnetic moments of a $u$ quark involved in those nuclei to an 
accuracy of $\sim 20\%$.
The exceptions are ${}^{113}_{~50}{\rm Sn}$ and ${}^{119}_{~52}{\rm Te}$ whose 
magnetic moments are consistent with the results of our calculation to within 
$50\div 90\%$.
${}^{113}_{~50}{\rm Sn}$ and ${}^{119}_{~52}{\rm Te}$ have respectively 
50 and 52 protons. 
50 is a magic number whereas 52 is not.
Other neutron-odd nuclei with magic numbers of protons
(${}^{15}_{~8}{\rm O}$ and ${}^{207}_{~82}{\rm Pb}$ in the state with $j=\frac12$,  
and ${}^{39}_{20}{\rm Ca}$, ${}^{57}_{28}{\rm Ni}$ and  ${}^{121}_{~50}{\rm Sn}$  in the state with 
$j= \frac32$) do not exhibit this discrepancy.   

Similar calculations can be done for proton-odd nuclei with $j= \frac12$, that 
is, with the use of Eq.~(\ref{mu-j+1_2}).
It is seen from Fig.~2(c) that the accuracy of $\sim 20\%$ between the 
results of our calculations and the data for the magnetic moments of such 
nuclei \cite{Fuller-} can be attained if we turn to the behavior of a $u$ quark.
Although it would be more physically reasonable to invoke the calculated 
magnetic moments of a $d$  quark for comparing them with those of proton-odd 
nuclei, the use of a $u$ quark for this purpose results in a better fit. 
This fact is rather puzzling and should be illuminated in the subsequent studies.  
\psfrag{A}[c][c][0.7]{$A$}
\psfrag{mu}[c][c][0.7]{$\mu/\mu_B$}
\psfrag{TheorymuoneQQQQQ}[r][r][0.6]{\footnotesize Theory $\mu_{A}=0.15\mu_B$}
\psfrag{TheorymutwoQQQQQ}[r][r][0.6]{\footnotesize Theory $\mu_{A}=0.20\mu_B$}
\psfrag{Theory}[r][r][0.6]{\footnotesize Theory}
\psfrag{Experiment}[r][r][0.6]{\footnotesize Experiment}

\begin{figure}[!ht]
\centerline{\includegraphics[height=\risheight,width=\riswidth,angle=270]{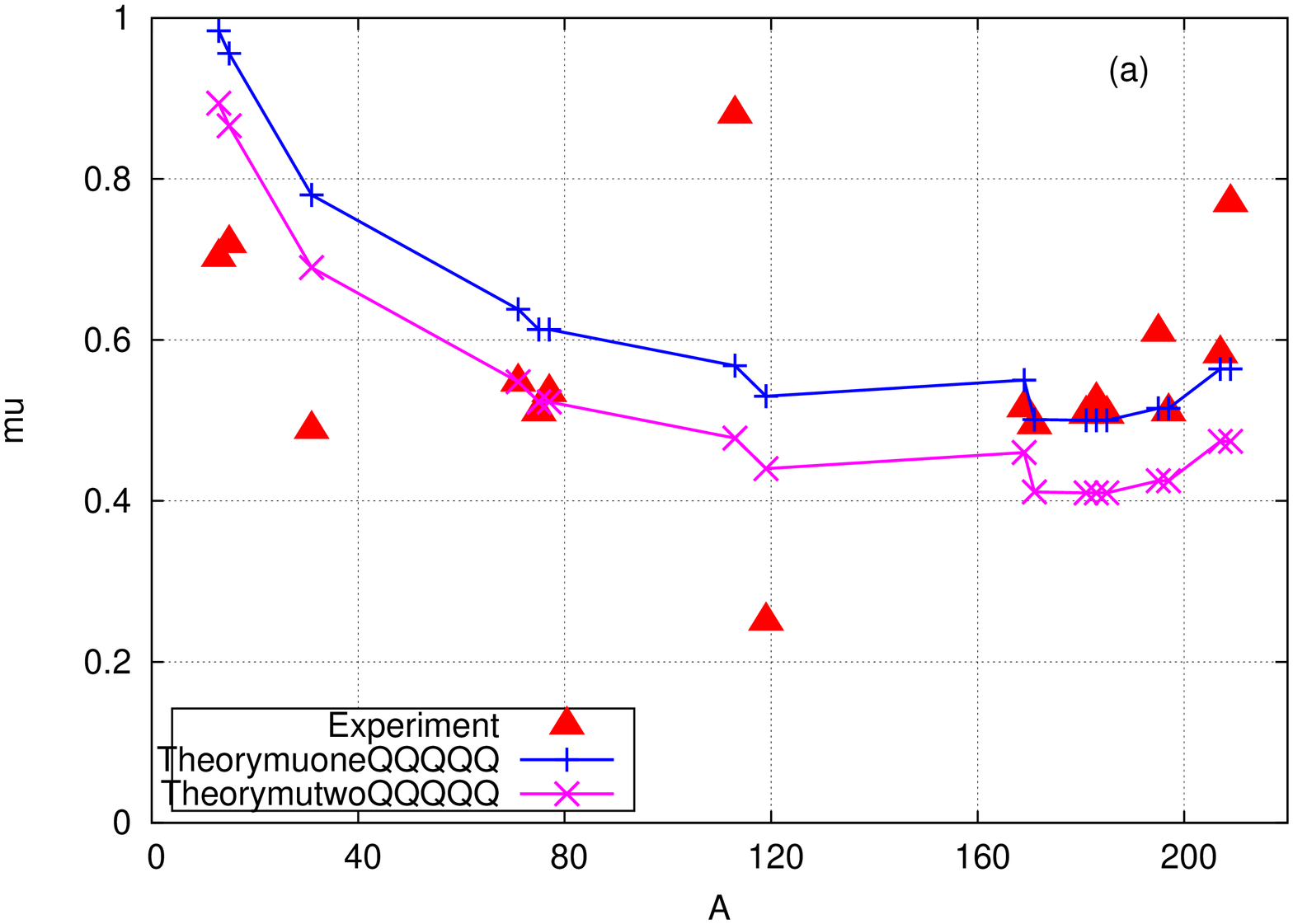}}
\centerline{\includegraphics[height=\risheight,width=\riswidth,angle=270]{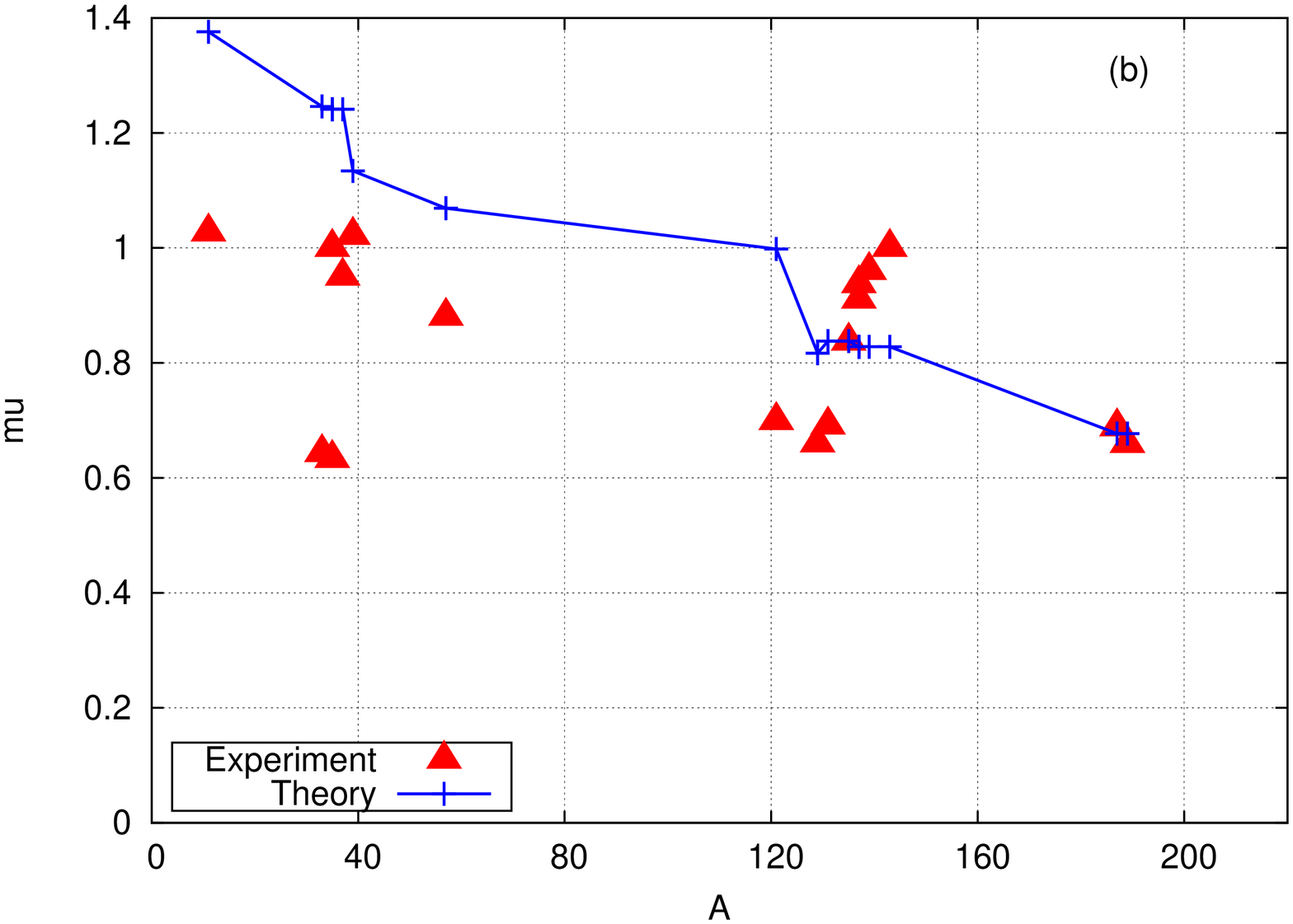}}
\centerline{\includegraphics[height=\risheight,width=\riswidth,angle=270]{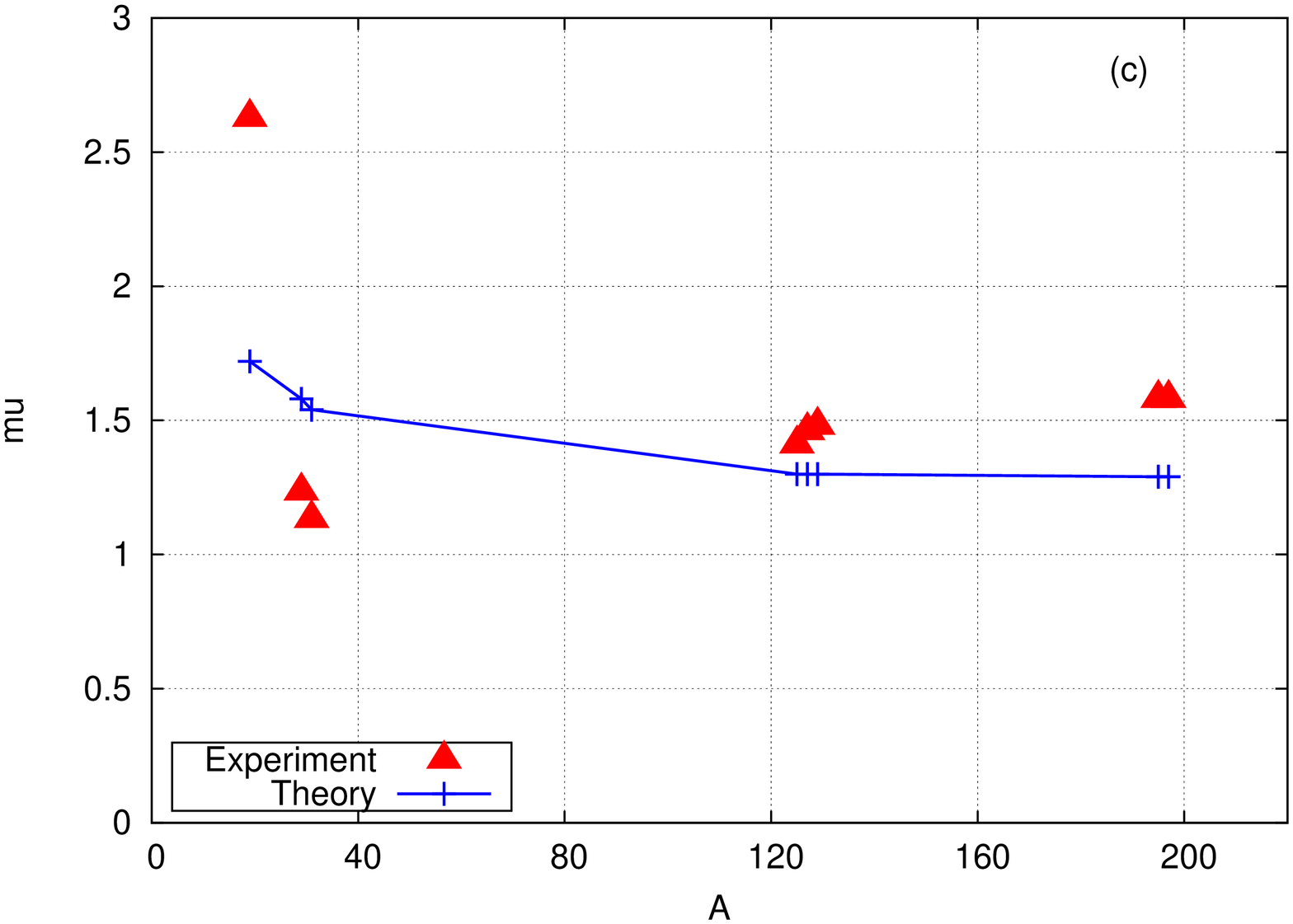}}
\caption{Comparison of the calculated magnetic moments with their experimental
values for (a) neutron-odd 
$(j=1/2)$, (b) neutron-odd $(j=3/2)$, (c) proton-odd $(j=1/2)$ nuclei}\label{nojthreehalfs}
\end{figure}

Since a single proton may be regarded as both a free hadron and the lightest 
nucleus,  ${}^{1}_{1}{\rm H}$, that is, a system subject to both spin- and 
pseudospin symmetry conditions, which are in conflict with each other, our 
argument is unsuitable for it.
In this connection, the idea of spherical singular cavity is to be applied to 
light nuclei (which fall within a transition region between the spin- and 
pseudospin symmetry coverages) with care.
The derivation of the properties of light nuclei in the framework of an 
effective theory to low-energy QCD is an intricate problem in itself.
The severity of the problem was recently manifested in Ref. \cite{Park} by 
giving theoretical evidence that six-quark color-neutral configurations 
compatible with the physical states of dibaryons cannot form stable bound
states.

\section{Conclusion}
Let us summarize the main results of this paper.
We consider the behavior of a single quark $Q$ contained in 
some nucleus, assuming that all remaining degrees of freedom of this 
nucleus were already integrated out, and the result of this integration yields 
a mean field which exerts on the quark $Q$.
We anticipate that the quark $Q$ is governed by the Lagrangian (\ref{QCD-Lagrangian}).
We invoke the semiclassical treatment, which implies that the extremal path 
contribution dominates the Feynman path integral.
Thus, the wave function of 
the quark $Q$ bears on a solution to the Dirac equation.
The quark $Q$ is affected by the Lorentz vector 
potential $A_\mu(x)$ and Lorentz scalar potential $\Phi(x)$ of the mean field.
We restrict our attention to spherically symmetric static interactions between 
the quark $Q$ and the mean field, and assume that the contribution of $A_\mu$ 
to the mean field is given by $A_0$. 
The pseudospin symmetry condition is imposed on $U_V=g_V A_0$ and $U_S=g_S\Phi$, 
namely $U_S=-U_V+C_s$.
We take the Cornell potential (\ref{Cornell}) as a phenomenological realization 
of both  $U_V$ and  $U_S$, and put $U_V=\frac12\,V_{\rm C}$.
This allows understanding of the fact that the interaction between the quark 
$Q$ and this mean field yields a spherical cavity of radius $r_{\rm sc}$ with
the boundary where the effective potential  $U(r;\varepsilon)$ is 
infinite. 
This keeps the quark $Q$ in this cavity from escaping.
We solve numerically  Eq.~(\ref{1D_Schroedinger}) with the parameters of the Cornell 
potential borrowed from the description of quarkonia, and obtain the energy levels 
$\varepsilon$ for  $\kappa=-1,-2$ and the corresponding sizes of the cavities
$r_{\rm sc}(\varepsilon)$.
The energy levels $\varepsilon_{n_r}$ turn out to be proportional to $\sqrt{n_r}$,
where ${n_r}$ is the radial quantum number determined by counting
the nodes of the radial amplitude $f_{n_r,\kappa}$.
We assume that ${n_r}$ equals the integral part of ${\cal A}^{2/3}$, where 
${\cal A}$ is the nucleon mass number.
This gives the relationship 
$r_{\rm sc}=R_0{\cal A}^{1/3}$, which is characteristic of the liquid drop 
model.
To verify that the assumption ${n_r}=[{\cal A}^{2/3}]$ is consistent with the
experimental data, we compare the magnetic dipole of the quark  $Q$ and that 
of the nucleus in which this quark is incorporated.
The agreement between the calculated and observed values of $\mu/{\mu}_{\rm B}$
is for the most part within $\sim 20\%$ which is better than expected when 
taken into account that the picture in which a single quark moving in a static 
spherically symmetric mean field applies to a rich variety of nuclei whose 
dynamical contents are highly tangled.

\vskip10mm
\noindent
{\Large{\bf Acknowledgements}}
\vskip5mm
\noindent
We thank S. Brodsky, M. Chaichian, E. Epelbaum, J. Ginocchio, 
T. Goldman, and K. Yazaki for fruitful discussions.
Critical remarks of an anonymous referee were of benefit to further clarification of
the present text.

\vskip10mm
\noindent
{\Large{\bf Appendix A}}
\vskip5mm
\noindent
We outline here the procedure of looking for numerical solutions to 
Eqs.~(\ref{Dirac_radia_f})--(\ref{B-df}) with imposing the conditions $U_S=U_V
=\frac12\,V_{\rm C}$.
Since our interest is with solutions that are regular at $r=0$, we use the 
ans\"atze $f(r)=u(r)r^{|\kappa|-1}$ and $g(r)=v(r)r^{|\kappa|-1}$.
The functions $u$ and $v$ satisfy the following integral equations
$$
u(r)=\frac{\kappa-|\kappa|}{\alpha_s}+(\varepsilon+m)\int_0^r ds\left[
\theta(-\kappa)+\theta(\kappa)\left(\frac{s}{r}\right)^{2|\kappa|}\right]v(s)\,,
\eqno(A.1)
$$
$$
v(r)=1+\int_0^r\! ds\!\left\{({\sigma}s-\varepsilon+m)\!\left[\theta(\kappa)+
\theta(-\kappa)\left(\frac{s}{r}\right)^{2|\kappa|}\right]\!u(s)+
\frac{\alpha_s(\varepsilon+m)}{2|\kappa|}\!
\left[\left(\frac{s}{r}\right)^{2|\kappa|}-1\right]\!v(s)\right\},
\eqno(A.2)
$$
where $\theta(\kappa)$ is the Heaviside step function. 
We write $u$ and $v$ as the Liouville--Neumann series,
$$
u=\sum_{k=0}^{\infty}u_k\,,\quad v=\sum_{k=0}^{\infty}v_k\,,
\eqno(A.3)
$$
where  $u_0=(\kappa-|\kappa|)/\alpha_s$, $v_0=1$,
$$
u_{k+1}=\Phi_{12}v_k\,,\quad 
v_{k+1}=\Phi_{21}u_k+\Phi_{22}v_k\,,\quad k\geq 0\,,
\eqno(A.4)
$$
$$
\Phi_{12}v(r)=(\varepsilon+m)\int_0^r ds\left[\theta(-\kappa)+
\theta(\kappa)\left(\frac{s}{r}\right)^{2|\kappa|}\right]v(s)\,,
\eqno(A.5)
$$
$$
\Phi_{21}u(r)=\int_0^r ds\,({\sigma}s-\varepsilon+m)\left[\theta(\kappa)+
\theta(-\kappa)\left(\frac{s}{r}\right)^{2|\kappa|}\right]u(s)\,,
\eqno(A.6)
$$
$$
\Phi_{22}v(r)=\frac{\alpha_s(\varepsilon+m)}{2|\kappa|}\int_0^r ds
\left[\left(\frac{s}{r}\right)^{2|\kappa|}-1\right]v(s)\,.
\eqno(A.7)
$$
Every term of these series is a polynomial. 
The coefficient of a given monomial can be found recursively taking into 
account that
$$
\Phi_{12}(r^k)=(\varepsilon+m)\left[\frac{\theta(-\kappa)}{k+1}+
\frac{\theta(\kappa)}{k+1+2|\kappa|}\right]r^{k+1}\,,
\eqno(A.8)
$$
$$
\Phi_{21}(r^k)=\sigma
\left[\frac{\theta(\kappa)}{k+2}+
\frac{\theta(-\kappa)}{k+2+2|\kappa|}\right]r^{k+2}-
(\varepsilon-m)\left[\frac{\theta(\kappa)}{k+1}+\frac{\theta(-\kappa)}{k+1+2|\kappa|}
\right]r^{k+1}\,,
\eqno(A.9)
$$
$$
\Phi_{22}(r^k)=-\frac{\alpha_s(\varepsilon+m)}{(k+1)(k+1+2|\kappa|)}\,r^{k+1}\,.
\eqno(A.10)
$$

\end{document}